\documentclass[twocolumn,fleqn,usenatbib,useAMS]{mnras}
\usepackage{lipsum}
\usepackage[switch]{lineno}
\usepackage{amssymb,amsmath}
\usepackage{graphicx}
\usepackage{booktabs}
\usepackage{hhline}
\usepackage{multirow}
\usepackage{color}
\usepackage{newtxtext,newtxmath}
\usepackage[T1]{fontenc}
\DeclareRobustCommand{\VAN}[3]{#2}
\let\VANthebibliography\thebibliography
\def\thebibliography{\DeclareRobustCommand{\VAN}[3]{##3}\VANthebibliography}

\newcommand{\degree}{^{\circ}}

\newcommand{\asca}{{\it ASCA\,}}
\newcommand{\rosat}{{\it ROSAT\,}}

\newcommand{\suzaku}{{\it Suzaku\,}}
\newcommand{\xmm}{{\it XMM\,}}
\newcommand{\xmmnewton}{{\it XMM-Newton\,}}

\newcommand{\swift}{{\it Swift\,}}

\newcommand{\nustar}{\textit{NuSTAR\,}}

\newenvironment{putjo}{\,\noindent\ignorespaces\fontfamily{pcr}\selectfont\ignorespaces}{\ignorespacesafterend\,}
\usepackage[symbol]{footmisc}

\title[Multi-epoch X-ray analysis of Mrk\,926]{Broadband spectral analysis of Mrk\,926 using multi-epoch X-ray observations}

\author[Chalise et al.]{
	S. Chalise,$^{1}$\thanks{E-mail: sulov.chalise@montana.edu}
	A. M. Lohfink,$^{1}$
	J. Chauhan,$^{1}$
	T. D. Russell,$^{2}$
	D. J. K. Buisson$^{3}$
	L. Mallick$^{4}$
	\\
	$^{1}$Department of of Physics, Montana State University, P.O. Box 173840, Bozeman, MT 59717-3840, USA\\
	$^{2}$INAF, Istituto di Astrofisica Spaziale e Fisica Cosmica, Via U. La Malfa 153, I-90146 Palermo, Italy\\
	$^{3}$Department of Astronomy, University of Southampton, Southampton SO17 1BJ, UK\\
	$^{4}$Cahill Center for Astronomy and Astrophysics, California Institute of Technology, Pasadena, CA 91125, USA
}

\date{Accepted XXX. Received YYY; in original form ZZZ}

\pubyear{2022}

\begin{document}
	\label{firstpage}
	\pagerange{\pageref{firstpage}--\pageref{lastpage}}
	\maketitle
	
	\begin{abstract}
		The X-ray spectra of some active galactic nuclei (AGN) show a soft X-ray excess, emission in excess to the extrapolated primary X-ray continuum below 2\,keV. Recent studies have shown that this soft excess can be described well as originating from either a relativistic ionized reflection, the extreme blurring of the reprocessed emission from the innermost region of the accretion disk, or Comptonization from an optically thick and warm region called the `warm corona', in which electron scattering is the dominant source of opacity. To constrain the origin of the soft excess in the Seyfert 1 galaxy Mrk\,926, we carry out an multi-epoch X-ray spectral study using observations from \textit{Suzaku} (2009), \textit{XMM-Newton} and \textit{NuSTAR} (2016), and \textit{NuSTAR} and \textit{Swift-XRT} (2021). The broadband X-ray spectra of Mrk\,926 contains: a thermally Comptonized primary continuum, a variable soft excess, and distant reflection. We find that in Mrk\,926 as in so many sources, it is difficult to make a definite statement as to what is causing the observed soft excess. A warm coronal-like component is slightly preferred by the data but a reflection origin is also possible. Using archival radio data, we detect an optically-thin radio component in our broadband study of Mrk\,926. While this component is consistent with an optically-thin radio jet, future multi-wavelength observations including high spatial resolution radio observations at multiple frequencies are required to probe the origin of the radio emission in more detail.
	\end{abstract}
	
	\begin{keywords}
		Mrk 926 -- MCG-02-28-22 -- Warm corona -- radiation mechanisms: non--thermal
	\end{keywords}
	\section{Introduction} \label{sec:intro}
	\par Active galactic nuclei (AGN) are powered by accretion onto the supermassive black hole (SMBH) at the galactic center of their host galaxies. They are highly energetic, and common X-ray sources in the universe. X-ray emission from AGN, produced at the innermost region, is a great probe to study the physical processes in the extreme gravity regime. A subgroup of AGN, the Seyfert 1 galaxies, are particularly useful, in an observational sense, due to the relatively unobscured view of their central engine.
	\par All Seyfert 1 X-ray spectra share key similarities. Their primary X-ray emission can be approximated as a cut-off powerlaw, and is consistent with being produced via thermal Comptonization of optical-UV photons from the accretion disk. This thermal Comptonization occurs in a compact region called the hot corona \citep{haardt1991,haardt1994}. A part of this primary emission can be reprocessed by interacting with different regions of the accretion disk or the dusty torus, generating a `reflection' spectrum. Reflection from distant matter generates a neutral Fe K$\alpha$ fluorescent emission line \citep{george1991,matt1991} at 6.4\,keV. If the reflection is from an area closer to the SMBH, Doppler shift and gravitational redshift broadens the fluorescent line profile \citep{fabian1989} and more complex line emission is produced as the material is ionized. The reflection spectrum is also accompanied by the ``Compton hump" feature around 20--30\,keV due to Compton scattering of the primary continuum from the disk or parsec (pc) scale torus \citep{ross1999}.
	\par In more than 50\,$\%$ of Seyfert 1 galaxies, soft X-ray emission in excess to that of the extrapolated hard X-ray continuum is observed below 2\,keV and referred to as `soft excess'. Its origin is still not completely understood, and is currently theorized to be either due to warm Comptonization or relativistic reflection from the accretion disk \citep{arnaud1985,walter1993,pico2005,winter2012,done2012}. The emitted X-rays can also be obscured by the gas and dust present in the line-of-sight, imparting absorption features. All these components make up a typical Seyfert X-ray spectrum. We can study the AGN phenomenon by analyzing this spectrum using physically motivated models.
	\par A group of nearby Seyfert 1 galaxies that show little to no X-ray absorption are the so-called ``bare Seyferts". Their study is free from the uncertainties introduced from the complex modeling of the absorption providing a clear measurement of the central region. Mrk\,926, also known as MCG-2-58-22, is a local (z=0.047) X-ray luminous ``bare" AGN classified in the optical as a Seyfert 1.5 galaxy. The lack of extended structures in the radio observations of Mrk\,926 (VLBA, 8\,GHz) with a high temperature sub-pc scale core supports its ``bare" classification \citep{mundell2000}. We usually expect cold and diffuse radio emissions from dusty obscurers. Also, Mrk\,926's 2--10\,keV X-ray flux is known to vary on a timescale of months to years \citep{choi2001,choi2002}. AGN variability can provide an additional probe into the scale and structure of the innermost region. Thus, a clear view of a variable central engine makes Mrk\,926 an interesting X-ray source.
	\par Mrk\,926 has been observed previously with many X-ray observatories such as \rosat, \asca, \xmm, and \textit{BeppoSAX} \citep[etc.]{weaver1995,choi2001,bianchi2004}. Those observations have revealed the presence of a variable soft excess \citep{ghosh1992}, a variable Fe K line profile \citep{weaver2001}, and a Compton hump \citep{bianchi2004} atop the primary continuum. This was also confirmed by the analysis of a 140\,ks \suzaku observation \citep{rivers}, which found a primary power-law continuum, a soft excess and placed tight constraints on the strength and nature of the reflection features. The authors also found no significant evidence for broadening of the Fe line expected from the inner accretion disk reflection \citep{rivers}. The same \suzaku and the 2016 joint \xmm-\nustar observations were studied by \citet{laha2021}, where the authors found the vanishing of Compton-hump between 2009 and 2016 observations with similar primary continuum. The authors proposed a scenario of a dynamic torus structure as an explanation \citep{laha2021}.
	\par A recent velocity-resolved reverberation mapping of Mrk\,926 was carried out by \citet{reverbmap} with a detailed spectroscopic and photometric variability study of its very broad emission lines (FWHM $\gg 4000$\,km\,s$^{-1}$). They found a drastic decrease of the optical continuum luminosity (50\% in 2.5 months) suggesting high variability. The inclination angle of the line emitting region was found to be $\sim$50$\degree$. The black hole mass was derived to be 1.1$(\pm{0.2})\times10^{8}$M$_{\odot}$, indicating a low Eddington ratio (3-8\%). Interestingly, they also found additional fast-response outer emission components, the Balmer satellites, which originate in a different, spatially distinct region such as a small-scale central radio jet.
	\par There are still several open questions regarding Mrk\,926's X-ray spectrum, mentioned in the following text. Using archival \rosat and \asca data, \citet{choi2001} found a broad iron line centered at $\sim$ 6.3\,keV. This is not unexpected from a bare nucleus as we have a clear view of the innermost emission regions where the relativistic broadening of the iron line happens. However, other studies such as \citet{rivers} (140\,ks \suzaku broadband data) and \citet{bianchi2004} (simultaneous $\sim 7$\,ks \xmm and \textit{BeppoSAX} broadband data) did not find a broadened line. Therefore, a pertinent question is whether the broad iron line is a transient phenomenon in Mrk\,926 or is its strength variable and is sometimes overwhelmed by the rest of the emission? If the iron line is variable, what is causing these changes? There is historical evidence of strong soft X-ray variability in Mrk\,926 \citep{ghosh1992}. The origin of the soft excess, and the physical process responsible for this variability is still unknown. In this study of Mrk\,926, we make an effort to answer these open questions.
	\par In this paper, we report the results from a 2009 \suzaku\,observation, a 2016 overlapping \xmm and \nustar campaign, and a 2021 \nustar and \swift-\textit{XRT} observation of Mrk\,926. We analyze these multi-epoch spectra and shed more light on the physical processes that govern the central engine of this AGN. The next section tabulates the new observations and describes the data reduction. We then perform a detailed spectral analysis in Section \ref{sec:span}. We subsequently discuss the results and their relevance for our understanding of Mrk\,926 and other Seyfert-like AGN in Section \ref{sec:discuss}. Finally, we summarize our findings in Section \ref{sec:conclusion}.
	\section{Observations and data reduction} \label{sec:dred}
	\subsection{Overview}
	In this paper, we analyze the broadband X-ray spectra of Mrk\,926 observed in three different epochs \citep[][this work]{rivers,laha2021}. The first epoch is the 2009 \suzaku observation, the second epoch is the joint 2016 \xmm-\nustar observation, and the third epoch is the 2021 \nustar \citep{Harrison2013} observation with a \swift-XRT pointing. The details of these epochs are provided in table \ref{obstable}.
	\begin{table}
		\centering
		\caption{The details of Mrk\,926 X-ray observations used in this work.}
		\label{obstable}
		\setlength{\tabcolsep}{0.7\tabcolsep}
		\begin{tabular}{|c|c|c|c|c|} 
			\hline\hline
			\textbf{Satellite}      & \textbf{ObsID}                            & \textbf{Date}                & \textbf{Net Exposure} & \textbf{Epoch}  \\ 
			\hline\hline
			\textit{Suzaku}                  & \textcolor[rgb]{0.2,0.2,0.2}{704032010}   & \textcolor[rgb]{0.2,0.2,0.2}{2009-12-02} & 139 ks                & 1               \\ 
			\hline
			\textit{XMM-Newton}              & \textcolor[rgb]{0.2,0.2,0.2}{0790640101}  & \textcolor[rgb]{0.2,0.2,0.2}{2016-11-21} & 59 ks                 & 2               \\ 
			\hline
			\multirow{2}{*}{\textit{NuSTAR}} & \textcolor[rgb]{0.2,0.2,0.2}{60201042002} & \textcolor[rgb]{0.2,0.2,0.2}{2016-11-21} & 106 ks                & 2               \\ 
			\cline{2-5}
			& 60761009002                               & 2021-07-04                               & 17 ks                 & 3               \\ 
			\hline
			\textit{Swift-XRT}               & 00089294001                               & 2021-07-04                               & 2 ks                  & 3               \\
			\hline
		\end{tabular}
	\end{table}
	\subsection{\textit{Suzaku}}
	The observation was performed by the XIS 0/1/3 CCD cameras and the Hard X-ray Detector (HXD) Positive Intrinsic Negative (PIN) silicone diodes. Our data reduction method is similar to \citet{rivers}, which also studied this observation. We generated the cleaned event files with the \suzaku calibration database released on 2018 October 10. The source spectrum was selected from a circular region of 170 arcsec radius, while the background was from a circular source-free region of 125 arcsec. The ARFs and RMFs were generated using \texttt{xisarfgen} and \texttt{xisrmfgen} tasks, respectively. The HXD-PIN data were reduced with \texttt{aepipeline} task. We utilized the "tuned" background event files to produce the non X-ray background spectra of HXD-PIN, to which the simulated cosmic X-ray background spectrum was added. 
	\subsection{\textit{XMM-Newton}}
	The \xmmnewton Observation data files (ODF) were processed using the \xmmnewton Science Analysis System (XMM-SAS v20.0). The MOS detector, operated in small window mode, is fully covered by the source photons and the selection of a source-free background region is not feasible. We thus focus our analysis on the EPIC-PN. The EPIC-PN data were first screened and periods of high particle backgrounds rejected. The spectra and light curves were created from the cleaned event files using \texttt{evselect}. We select an annular source region of 40 arcsec outer radius and 8 arcsec inner radius in order to mitigate mild pile-up. We also select a circular background region of 60 arcsec. Responses were created with \texttt{arfgen} and \texttt{rmfgen} tools. In \texttt{arfgen}, we apply the empirical correction of the EPIC effective area with the parameter \texttt{applyabsfluxcorr=yes}. This correction is essential to rectify the spectral shape mismatch seen with the simultaneous \nustar observation. 
	\subsection{\nustar}
	The \nustar data from both detectors, FPMA and FPMB, were reduced using the standard pipeline (\begin{putjo}NUPIPELINE\end{putjo}) of the \nustar Data Analysis Software (\begin{putjo}NUSTARDAS\end{putjo}v2.1.1). Calibration files from \nustar \begin{putjo}CALDB\end{putjo} v20210202 were used. We utilized the background filtering reports provided by the \textit{NuSTAR} team to reduce the effect of high particle background and produce cleaned event files. We selected SAAMODE ``optimized" and exclude the ``tentacle" region minimizing the effect of the South Atlantic Anomaly (SAA) and optimizing exposure time and background level. We then extracted the spectra and light curves from a circular source region of 40 arcsec and a circular background region of 100 arcsec close to the source.
	\subsection{\textit{Swift-XRT}}
	The \swift\textit{-XRT} spectra were obtained using the \swift\textit{-XRT} data products generator \citep{evans2009} supplied by the UK Swift Science Data Centre at the University of Leicester.
	\section{Spectral analysis} \label{sec:span}
	\par
\begin{figure}
		\centering 
		\includegraphics[width=0.48\textwidth]{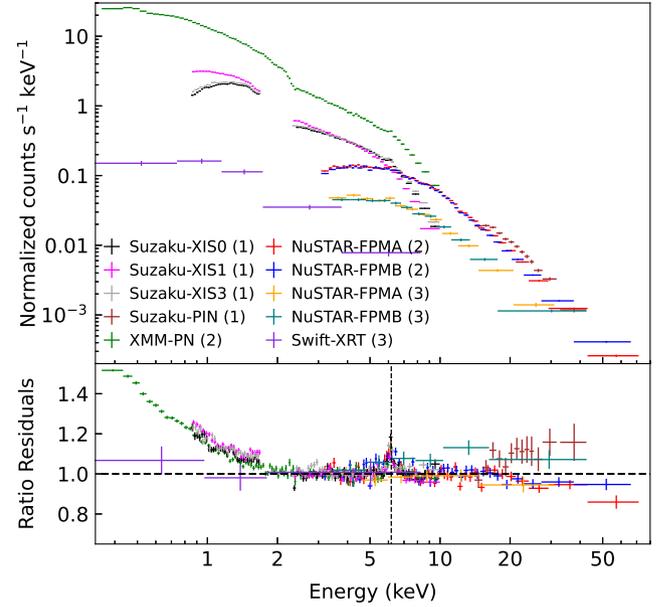}
		\caption{\textbf{Top:} All counts spectra for \textit{Suzaku}, \textit{XMM}, \textit{NuSTAR} and \textit{Swift-XRT} are shown; the spectra have been binned further for plotting. The number in brackets denote their respective epoch. \textbf{Bottom:} The ratio residuals after fitting the spectra with a power-law in the range of 3.0--5.0\,keV. The dashed line indicates the neutral Fe-K$\alpha$ energy.}
		\label{alldata}
	\end{figure}

 In section \ref{sec:dred}, we generated the multi-epoch X-ray spectra of Mrk\,926 using \suzaku\,(2009), joint \xmm-\nustar\,(2016) and joint \nustar-\swift\,(2021) observations. We bin them such that the detector resolution is over-sampled by a factor of 3, ensuring the minimum signal to noise ratio of 10, except for the \swift-XRT spectra which were binned to a minimum of 20 counts per bin. We use the energy range of 0.85--1.7\,keV, 2.3--10.0\,keV for XIS; 13.0--45.0\,keV for PIN; 0.3--10.0\,keV for PN; 3.0--75.0\,keV for FPM; 0.3--10.0\,keV for XRT. The resulting spectra are plotted in the top subplot of Figure \ref{alldata} with further binning for clarity. We carry out the spectral analysis and model fitting with \textit{XSPEC} 12.11.1 \citep{Arnaud1996} using the chi-square minimization technique; all uncertainties are reported at 90\% confidence.
	\par In addition, our models always include: a cross correlation constant that accounts for any cross-calibration flux offsets among the different spectra in the same epoch; a galactic absorption component that accounts for the ISM absorption from our galaxy using the ISM absorption model \texttt{TBabs} \citep{Wilms2000} with a column density ($N_{\text{H}} = 2.86\times 10^{20}$ cm$^{-2}$) obtained from the N$_{\text{H}}$ Tool \citep{Kalberla2005} using the latest HI4PI data. The cross-calibration constant for \suzaku-PIN is fixed at 1.16 which is the expected value for XIS-nominal pointing. 
	\par A power-law spectrum is a rough description of an AGN primary continuum in the X-ray band. Therefore to explore the additional X-ray components, We initially fit a power-law to each epoch in the energy range 3--5\,keV, where the spectra are likely dominated by the primary continuum. We then apply this model, without fitting, to the full energy range and search for residual features (Figure \ref{alldata}, bottom). The three epochs show variations in both low and hard end of the X-ray spectrum.
	\par We notice smoothly-rising positive residuals towards the soft energies ($<$2\,keV) indicative of a soft excess for Epoch 1 and 2. This could also be present in Epoch 3 but simply not resolved by \textit{Swift-XRT}. We also observe a line-like feature at the neutral Fe-K$\alpha$ energy. At higher energies ($>$15\,keV), Epoch 2 shows a clear excess. These features can be associated with an X-ray reflection spectrum, which are the reprocessed hard X-ray photons from the hot corona after being backscattered off the AGN accretion disk or a distant reflector. This reflection spectrum contains two main features: the 6.4\,keV Fe-K$\alpha$ line along with a bump-like feature that usually peaks around 25\,keV. If this reflection occurs close to the black hole, these features can be relativistically smeared. In some cases, this relativistic reflection can explain the observed soft-excess \citep{garcia2019}. Thus, we check if the reflection scenario can adequately explain the broadband multi-epoch spectra of Mrk\,926.
	\subsection{Dual reflection}\label{sec:dualrefl}
	\par AGNs are observed to be intrinsically variable in short timescales of a week or less. So in each epoch, we let the primary continuum to be different. However, we assume a stable reflecting medium between epochs as their variability timescale is generally much larger.
	\par We model the primary continuum using the thermal Comptonization model \begin{putjo}Nthcomp\end{putjo} \citep{nthcomp}. The powerlaw photon index (\texttt{$\Gamma$}) and its normalization is allowed to vary among different epochs. The disk-blackbody seed photon temperature (\texttt{kT$_{bb}$}) is not sensitive to the fit and is fixed at 0.05\,keV. The electron temperature (\texttt{kTe}) is a free parameter linked among epochs. For reflection off this primary continuum, we use \begin{putjo}xillverCp\end{putjo} \citep{Garcia2013} for the distant reflection from the outer disk and torus, and \begin{putjo}relxillCp\end{putjo} \citep{Garcia2014} for relativistic reflection. The reflection fraction for both are negative (reflection spectrum only) and variable among epochs while the Power-law photon index (\texttt{$\Gamma$}), normalization and Electron temperature (\texttt{kTe}) are linked to the primary continuum of the respective epoch. The inclination angle (\texttt{incl}) of both reflectors is linked and only allowed to vary between the range of 45$\degree$--55$\degree$. This inclination angle range was determined by \citet{reverbmap} using high-cadence spectroscopic variability campaign of Mrk\,926 in the optical band. The ionization parameter (\texttt{logxi}) of the distant reflector is fixed at 0 while it is a free parameter for the relativistic reflector. The iron abundance (\texttt{A$_{\text{fe}}$}) for both reflectors are free but not allowed to vary among epochs. For the relativistic reflection: the inner (\texttt{R$_{\text{in}}$}) and outer (\texttt{R$_{\text{out}}$}) radius of the accretion disk are fixed at the innermost stable circular orbit (ISCO) and 400 R$_{\text{g}}$ (gravitational radii) respectively; the emissivity index up to 15 gravitational radii is a variable parameter among epochs while the emissivity index beyond that is a free parameter linked among epochs; the spin of the black hole (\texttt{a}) is fixed at the maximum value of 0.998.
	\par Due to the limited photon counts, we find some reflector parameters to be unconstrained for Epoch 3. Thus, we link all \begin{putjo}relxillCp\end{putjo} parameters except the reflection fraction from Epoch 3 to Epoch 1. The fit for this dual reflection scenario is satisfactory with a chi-square of 1582 for 1143 degrees of freedom. We constrained the electron temperature (\texttt{kTe}) to $>284$\,keV; the iron abundance of the distant reflector to 5.0$\substack{+2.3 \\ -0.3}$ times the solar value; the iron abundance of the relativistic reflector to 1.6$\pm{0.4}$ times the solar value; the ionization parameter (\texttt{logxi}) to 2.70$\pm{0.02}$. The detailed parameter values of the fit are shown in Table \ref{dualrefltable}. We note that the ratio residuals plot of this fit (Figure \ref{ratio_nth_relxill}, top) shows that it cannot explain the data above 30\,keV perfectly. 
	\subsection{Warm corona}\label{sec:dualcorona}
	\par The soft-excess can also be modeled as a Comptonized emission from a separate component, often thought to be a warm and optically thick region called a warm corona \citep{petrucci2018}. We investigate if such a warm corona-like component instead of a relativistic reflection can fit the multi-epoch spectra.
	\par
\begin{figure}
		\centering 
		\includegraphics[width=0.48\textwidth]{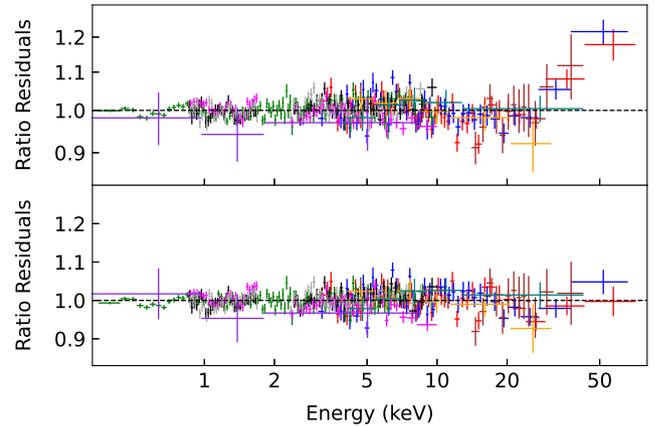}
		\caption{The top panel shows the ratio residuals for the dual reflection model fit in Section \ref{sec:dualrefl}; the bottom panel shows the ratio residuals for the warm corona model fit in Section \ref{sec:dualcorona}. The colors represent the same spectra as Figure \ref{alldata}. }
		\label{ratio_nth_relxill}
	\end{figure}

 We retain the same model structure for the primary continuum and its distant reflection, and replace the relativistic reflection with a warm corona. We model the warm corona using the thermal Comptonization model \begin{putjo}Nthcomp\end{putjo} \citep{nthcomp}. The Power-law photon index (\texttt{$\Gamma$}), the electron temperature (\texttt{kTe}) and the normalization for warm corona is allowed to vary among different epochs. The disk-blackbody seed photon temperature (\texttt{kT$_{bb}$}) is fixed at 0.05\,keV. Similar to Section \ref{sec:dualrefl}, we find the warm coronal parameters to be unconstrained for Epoch 3. Thus, we link all except the normalization parameter from Epoch 3 to Epoch 1. This model fits data well with a chi-square of 1479 for 1139 degrees of freedom. We constrained the electron temperature of the hot corona to 39$\pm{9}$\,keV; the iron abundance of the non-relativistic reflector to $>8.4$ times the solar value. The detailed parameter values of the fit are shown in Table \ref{warmcoronatable}. The ratio residuals plot of this fit (Figure \ref{ratio_nth_relxill}, bottom) shows that this model explains the data above 30\,keV better than the dual reflection scenario. We note that this fit is slightly better with the inclusion of additional narrow Gaussian lines in the \xmm\, spectra, but for simplicity these were omitted as the results are unaffected. 
	\begin{table*}
		\centering
		\caption{Spectral parameters for the dual reflection model (Section \ref{sec:dualrefl}) containing the primary continuum, distant reflection and relativistic reflection fitted to the multi-epoch data. L1 denotes that the parameter is linked to the respective Epoch 1 parameter whereas\, $f$ means the parameter is fixed at the given value.}
		\label{dualrefltable}
		\begin{tabular}{|c|c|l|l|l|l|} 
			\hline
			\textbf{Model~}                                                                                    & \multicolumn{1}{l|}{\textbf{Parameter description}} & \textbf{Symbol (Units)}    & \multicolumn{1}{c|}{\textbf{Epoch 1}} & \multicolumn{1}{c|}{\textbf{Epoch 2}} & \multicolumn{1}{c|}{\textbf{Epoch 3}}  \\ 
			\hline
			\multirow{3}{*}{\begin{tabular}[c]{@{}c@{}}Primary\\Continuum\\(nthcomp)\end{tabular}}             & Photon Index                                        & \texttt{$\Gamma$}                      & \multicolumn{1}{c|}{1.83$\pm{0.06}$}                 & \multicolumn{1}{c|}{1.88$\pm{0.05}$}                 & \multicolumn{1}{c|}{1.73$\pm{0.04}$}                  \\ 
			\cline{2-6}
			& Electron Temperature                                & \texttt{kTe} (keV)                & \multicolumn{1}{c|}{$>$284}                 & \multicolumn{1}{c|}{L1}                 & \multicolumn{1}{c|}{L1}                  \\ 
			\cline{2-6}
			& Normalization                                       & \texttt{norm} (10$^{-3}$)                & \multicolumn{1}{c|}{14.1$\pm{0.2}$}                 & \multicolumn{1}{c|}{12.2$\pm{0.2}$}                 & \multicolumn{1}{c|}{4.8$\pm{0.3}$}                  \\ 
			\hline
			\multirow{4}{*}{\begin{tabular}[c]{@{}c@{}}Distant\\Reflection\\(xillverCp)\end{tabular}} & Iron Abundance                                      & \texttt{Afe} (solar)                & \multicolumn{1}{c|}{5.0$\substack{+2.3 \\ -0.3}$}                 & \multicolumn{1}{c|}{L1}                 & \multicolumn{1}{c|}{L1}                  \\ 
			\cline{2-6}
			& Ionization parameter                                & \texttt{logxi} (ergs\,cm$^-2$\,s$^{-1}$)         &     		\multicolumn{1}{c|}{0$^{f}$}                                  &   \multicolumn{1}{c|}{L1}                                    &          \multicolumn{1}{c|}{L1}                              \\ 
			\cline{2-6}
			& Inclination                                         & \texttt{incl} ($\degree$)              & \multicolumn{1}{c|}{53$\pm{2}$}                 & \multicolumn{1}{c|}{L1}                 & \multicolumn{1}{c|}{L1}                  \\ 
			\cline{2-6}
			& Reflection fraction                                 & \texttt{refl\_frac} (-10$^{-3}$)         &   \multicolumn{1}{c|}{3.0$\pm{0.4}$}                                &       \multicolumn{1}{c|}{5.5$\pm{0.6}$}                                &            \multicolumn{1}{c|}{7.4$\pm{3.5}$}                            \\ 
			\hline
			\multirow{10}{*}{\begin{tabular}[c]{@{}c@{}}Relativistic\\Reflection\\(relxillCp)\end{tabular}}    & Inner emissivity index                              & \texttt{Index1}                     & \multicolumn{1}{c|}{4.5$\pm{0.3}$}                                      & \multicolumn{1}{c|}{$>$9.8}                                      & \multicolumn{1}{c|}{L1}                                       \\ 
			\cline{2-6}
			& Outer emissivity index                              & \texttt{Index2}                     & \multicolumn{1}{c|}{$>$4.0}                                      & \multicolumn{1}{c|}{L1}                                      &        \multicolumn{1}{c|}{L1}                                \\ 
			\cline{2-6}
			& Inner-outer break                                   & \texttt{Rbr} (gravitational radii)  & \multicolumn{1}{c|}{15$^{f}$}                                      & \multicolumn{1}{c|}{L1}                                      & \multicolumn{1}{c|}{L1}                                       \\ 
			\cline{2-6}
			& Spin of the black hole                              & \texttt{a}                          & \multicolumn{1}{c|}{0.998$^{f}$}                                      & \multicolumn{1}{c|}{L1}                                      & \multicolumn{1}{c|}{L1}                                       \\ 
			\cline{2-6}
			& Inner radius                                        & \texttt{Rin} (ISCO)                 & \multicolumn{1}{c|}{1$^{f}$}                                      & \multicolumn{1}{c|}{L1}                                      & \multicolumn{1}{c|}{L1}                                       \\ 
			\cline{2-6}
			& Outer radius                                        & \texttt{Rout} (gravitational radii) & \multicolumn{1}{c|}{400$^{f}$}                                      & \multicolumn{1}{c|}{L1}                                      & \multicolumn{1}{c|}{L1}                                       \\ 
			\cline{2-6}
			& Ionization parameter                                & \texttt{logxi} (erg\,cm$^{-2}$\,s$^{-1}$)         &  \multicolumn{1}{c|}{2.70$\pm{0.02}$}                                     & \multicolumn{1}{c|}{L1}                                      & \multicolumn{1}{c|}{L1}                                       \\ 
			\cline{2-6}
			& Iron Abundance                                      & \texttt{Afe} (solar)                & \multicolumn{1}{c|}{1.6$\pm{0.4}$}                                      & \multicolumn{1}{c|}{L1}                                      & \multicolumn{1}{c|}{L1}                                       \\ 
			\cline{2-6}
			& Reflection fraction                                 & \texttt{refl\_frac} (-10$^{-3}$)         & \multicolumn{1}{c|}{6.1$\pm{0.5}$}                                      & \multicolumn{1}{c|}{8.4$\pm{0.3}$}                                      & \multicolumn{1}{c|}{$>$6.4}                                       \\
			\hline
		\end{tabular}
	\end{table*}
	\begin{table*}
		\centering
		\caption{Spectral parameters for the warm corona model (Section \ref{sec:dualcorona}) the primary continuum, distant reflection and a warm corona fitted to the multi-epoch data. L1 denotes that the parameter is linked to the respective Epoch 1 parameter whereas \,$f$ means the parameter is fixed at the given value.}
		\label{warmcoronatable}
		\begin{tabular}{|c|c|l|l|l|l|} 
			\hline
			\textbf{Model~}                                                                                    & \multicolumn{1}{l|}{\textbf{Parameter description}} & \textbf{Symbol (Units)}    & \multicolumn{1}{c|}{\textbf{Epoch 1}} & \multicolumn{1}{c|}{\textbf{Epoch 2}} & \multicolumn{1}{c|}{\textbf{Epoch 3}}  \\ 
			\hline
			\multirow{3}{*}{\begin{tabular}[c]{@{}c@{}}Primary\\Continuum\\(nthcomp)\end{tabular}}             & Photon Index                                        & \texttt{$\Gamma$}                      & \multicolumn{1}{c|}{1.69$\pm{0.01}$}                 & \multicolumn{1}{c|}{1.77$\pm{0.01}$}                 & \multicolumn{1}{c|}{1.70$\pm{0.02}$}                  \\ 
			\cline{2-6}
			& Electron Temperature                                & \texttt{kTe} (keV)                & \multicolumn{1}{c|}{39$\pm{9}$}                 & \multicolumn{1}{c|}{L1}                 & \multicolumn{1}{c|}{L1}                  \\ 
			\cline{2-6}
			& Normalization                                       & \texttt{norm} (10$^{-3}$)                & \multicolumn{1}{c|}{12.7$\pm{0.5}$}                 & \multicolumn{1}{c|}{13.1$\pm{0.1}$}                 & \multicolumn{1}{c|}{4.8$\pm{0.2}$}                  \\ 
			\hline
			\multirow{4}{*}{\begin{tabular}[c]{@{}c@{}}Distant\\Reflection\\(xillverCp)\end{tabular}} & Iron Abundance                                      & \texttt{Afe} (solar)                & \multicolumn{1}{c|}{$>$8.4}                 & \multicolumn{1}{c|}{L1}                 & \multicolumn{1}{c|}{L1}                  \\ 
			\cline{2-6}
			& Ionization parameter                                & \texttt{logxi} (ergs\,cm$^-2$\,s$^{-1}$)         &     		\multicolumn{1}{c|}{0$^{f}$}                                  &   \multicolumn{1}{c|}{L1}                                    &          \multicolumn{1}{c|}{L1}                              \\ 
			\cline{2-6}
			& Inclination                                         & \texttt{incl} ($\degree$)              & \multicolumn{1}{c|}{$<$55}                 & \multicolumn{1}{c|}{L1}                 & \multicolumn{1}{c|}{L1}                  \\ 
			\cline{2-6}
			& Reflection fraction                                 & \texttt{refl\_frac} (-10$^{-3}$)         &   \multicolumn{1}{c|}{2.6$\pm{0.4}$}                                &       \multicolumn{1}{c|}{2.3$\pm{0.4}$}                                &            \multicolumn{1}{c|}{4.6$\pm{2.2}$}                            \\ 
			\hline
			\multirow{3}{*}{\begin{tabular}[c]{@{}c@{}}Warm Corona\\(nthcomp)\end{tabular}}                    & Photon Index                                        & \texttt{$\Gamma$}                   & \multicolumn{1}{c|}{2.84$\pm{0.1}$} & \multicolumn{1}{c|}{2.63$\pm{0.1}$} & \multicolumn{1}{c|}{L1}  \\ 
			\cline{2-6}
			& Electron Temperature                                & \texttt{kTe} (keV) & \multicolumn{1}{l|}{$>$0.85} & \multicolumn{1}{l|}{0.23$\pm{0.02}$} &
			\multicolumn{1}{l|}{\,\,\,\,\,\,\,\,\,L1}  \\ 
			\cline{2-6}
			& Normalization~                                      & \texttt{norm} (10$^{-3}$)             & \multicolumn{1}{l|}{4.3$\pm{0.6}$} & \multicolumn{1}{l|}{1.5$\pm{0.2}$} & \multicolumn{1}{l|}{$<$0.16}  \\
			\hline
		\end{tabular}
	\end{table*}
	\subsection{Mrk\,926 at Lower Frequencies} \label{sec:cjet}
	Using the Karl G. Jansky Very Large Array (VLA) observations at 4.8\,GHz with arcsec resolution, \citet{vla} found Mrk\,926 to be a relatively powerful radio source for a type 1 Seyfert with an unresolved core. \citet{mundell2000} studied the high angular resolution ($\sim$2\,mas) radio continuum observations of Mrk\,926 using the Very Long Baseline Array (VLBA) at 8.4\,GHz. They found the core to be unresolved with brightness temperature in excess of 10$^8$\,K and size less than 1\,pc. They found the radio core to be consistent with a non-thermal synchrotron self-absorption region without a Doppler boost. Mrk\,926 was also monitored at 22\,GHz with 1 arcsec resolution by the VLA \citep{jvla}. Using these past radio observations at arcsec or sub-arcsec resolution and additional archival flux measurements at other bands from Far-IR up to X-rays, we study the nature of the self-absorbed synchrotron core.
	\par
\begin{figure}
		\centering 
		\includegraphics[width=0.48\textwidth]{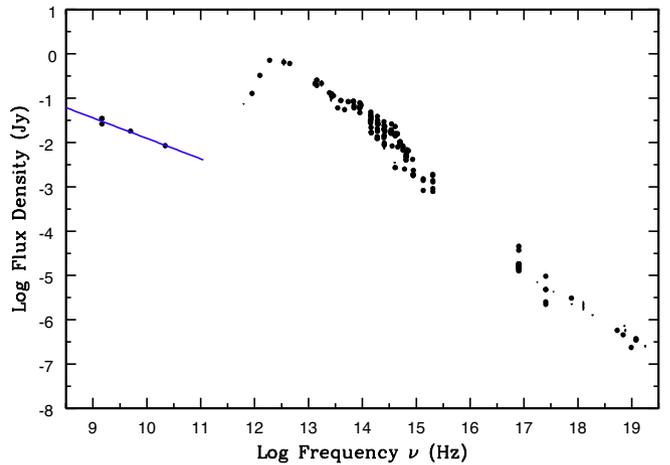}
		\caption{The SED generated by the online SED builder tool$^{1}$ for Mrk\,926 using archival flux measurements. The blue line is a power-law fit to the arcsec or higher resolution radio data.}
		\label{synchrotron}
	\end{figure}

 We use the online Spectral SED builder tool\footnote[1]{The SED builder is an online tool dedicated to multi-frequency data visualization, together with fitting routines useful for extracting refined scientific products. Provided by the Space Science Data Center (SSDC): http://www.ssdc.asi.it} to generate an spectral energy distribution (SED) using archival flux measurements of Mrk\,926. The SED is shown in Figure \ref{synchrotron}. We find the radio SED to be consistent with a power-law with a steep spectral slope (F$_{\nu} \propto \nu^{-0.54}$) which is indicative of an optically-thin component \citep{eckart1986}.
	\par In the optical reverberation mapping of Mrk\,926 \citep{reverbmap}, a Balmer satellite component, consistent with the presence of a compact jet, was found. A compact jet is continually-accelerated along the jet, where we detect single synchrotron spectra from different distances along the jet resulting in a flat to slightly inverted radio spectrum at GHz frequencies \citep{Blandford1979,Panessa2019}. The flat/inverted radio spectrum will break at some higher frequency $\nu_{b}$, either at the particle acceleration region in a shocked zone or the base of the jet, into a steep spectrum \citep{Blandford1979,Koljonen2015}.
	\par There could be an underlying compact jet in Mrk\,926 with the break frequency present at MHz or higher frequencies. This jet structure, if present, could be identified at higher spatial resolutions. However, with the available archival data, we cannot postulate the origin of this radio emission. A future multi-wavelength observing campaign of Mrk\,926 including higher spatial resolution radio observations at multiple frequencies is highly desired to study it in detail. This will allow us to perform a full SED modeling including the disk, host-galaxy etc. without the complications from variability, and resolve this structure.
	\section{Discussion} \label{sec:discuss}
	\subsection{Comparison with results from previous X-ray studies} \label{sec:prevstud}
	\par
\begin{figure}
		\centering 
		\includegraphics[width=0.48\textwidth]{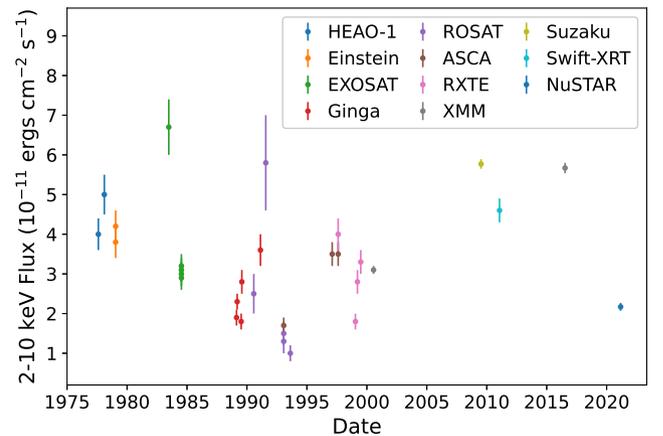}
		\caption{Historic 2--10\,keV flux variation plot adapted from \citet{choi2002} and updated with recent values.}
		\label{hist-flx}
	\end{figure}

 Historically, Mrk\,926 has been observed to show flux variations on longer timescales (months and years). Using data from various X-ray satellites covering more than 20 years (1977--1999), \citet{choi2002} reported the X-ray continuum flux variations in Mrk\,926. We update their observed 2-10\,keV X-ray flux variations plot (see Figure \ref{hist-flx}). We find that Mrk\,926 still shows significant continuum variability.
	\par Additionally, Mrk\,926 also shows variations in its iron line profile \citep{weaver2001,bianchi2004}. \citet{choi2001}, using ROSAT and ASCA archival data, found the line profile dominated by a broad component. \citet{bianchi2004}, using a $\sim$7\,ks \xmm and a simultaneous BeppoSAX observation, found the presence of a Fe K$\alpha$ and a Fe XXVI line. Meanwhile \citet{rivers}, using Suzaku observation, found the need for Fe K$\alpha$, K$\beta$, and an Fe K shell absorption edge to explain the observed line profile, and concluded that a broad component, if present, must be weak.
	\par The 2009 \textit{Suzaku} observation (Epoch 1) were described by \citet{rivers} using a torus model and an additional power-law component for soft excess. The torus was found to have subsolar iron abundance with a column density, \texttt{log(N$_{\texttt{H}}$)} of 24.55$\pm{0.15}$ while the inclination angle and the covering factor was fixed to 30$\degree$ and 0.5 respectively. The same observation was used by \citet{noda2013} to study the nature of soft excess in Mrk\,926 using intensity-correlated spectral analysis. They found that the thermal Comptonization interpretation can explain the broadband \suzaku data including the soft excess, while the relativistic reflection interpretation was failed to explain it. This is in agreement with our results in Section \ref{sec:dualrefl} and \ref{sec:dualcorona}.
	\par The 2009 \textit{Suzaku} and 2016 \xmm-\nustar\, observations (Epoch 1 and 2) were studied by \citet{laha2021}. They modeled the soft excess with just the relativistic reflection and found that between two epochs: the primary continuum is constant both in photon index and luminosity; the soft excess flux decreases by a factor of 2; the Compton hump vanishes; the narrow Fe-k$\alpha$ line becomes marginally broad and its flux doubles. They argue since the primary continuum does not change but the Compton hump vanishes between two epochs, the reprocessing medium must be changing. They speculate that this could be explained by a dynamic torus where the equatorial toroidal disk produces the narrow Fe-k$\alpha$ emission while the outflowing component generates the variable Compton hump.
	\par To explore this hypothesis, we replace \begin{putjo}xillverCp\end{putjo}  with a torus reprocessor model \begin{putjo}borus12\end{putjo}\citep{borus2018} in our multi-epoch analysis. Assuming either a stable or a variable torus structure between epochs, we find minimal changes to the fit and still see a small $>30$\,keV excess in the dual reflection scenario. For the warm corona scenario, we can get a fit similar to our best fit with a stable torus reprocessor. Additionally, recent simulations show that the outflowing component of the torus can change the equivalent width of some X-ray fluorescent lines but does not significantly influence the X-ray spectra \citep{polarsim}. Thus, we deem the scenario presented in \citet{laha2021} to be unlikely considering the warm corona/dual reflection model with a distant reflector, either torus or an accretion disk can describe the multi-epoch broadband X-ray spectra of Mrk\,926 without requiring significant changes to the reprocessor between epochs. \\
	\vspace{-20pt}
	\subsection{Soft excess Origin}
	\label{disc:softexcess}
	\par Recent studies debate two possible origins of AGN soft excess: the relativistic ionized reflection of the continuum from the inner accretion disk, and the Comptonization of the accretion disk photons in an optically thick, warm corona \citep[e.g.][]{petrucci2013,garcia2019}.
	\par There are many AGN for which the relativistic reflection origin of soft excess works well \citep[e.g.][]{mallick2018,jiang2019,middei2020}. In this scenario, the extreme gravitational blurring blends the rich forest of fluorescent emission lines produced by the reprocessing of X-ray photons in the innermost regions of accretion disk creating a smooth soft excess. However, this scenario requires extreme parameters, such as maximum black hole spin, a very low and compact hot corona, and a very high density for the inner accretion disk \citep{garcia2019}. From Section \ref{sec:span}, we find that the relativistic reflection struggles to describe the broadband X-ray spectra of Mrk\,926 well above 30\,keV. This could either mean that the soft excess in Mrk\,926 is not due to the relativistic reflection alone or that our simplistic reflection modeling still has shortcomings.
	\par Testing the warm coronal origin of soft excess in a sample of 22 unabsorbed radio-quiet AGN with combined 100 observations, \citet{petrucci2018} found it to be a good description to more than 90\% of the sample with the warm coronal temperature in the range $\sim 0.1-1$\,keV, and the optical depth in $\sim 10-40$ range. However, at these temperatures and optical depth, the atomic opacities is likely to dominate over the Thomson opacities and create strong absorption features instead of a smooth soft excess \citep{garcia2019}. This would also lead to rapid cooling of the warm corona, requiring a well-tuned heating process that constantly offsets this \citep{ballantyne2020a,ballantyne2020b}. \citet{petrucci2020} studied the radiative equilibrium and emission of a warm corona in the presence of an internal heating source and found that if this corona is covering a large part of a weakly dissipative accretion disk, it can generate emission matching the observed soft-excess. One such source of internal heating could be the presence of magnetized accretion flow.
	\par From Section \ref{sec:span}, we find that the warm corona model can model the broadband X-ray spectra of Mrk\,926 in all three epochs well at all energies. Thus, it is certainly a possibility for the soft excess origin in this source. To make a strong statement about the nature of the soft excess in this source, high quality data covering the 50-100\,keV range would be needed. Finally, we note that since we are unable to constrain the nature of radio emitter at high frequencies, we cannot fully exclude any flux contribution at soft X-rays. 
	\section{Summary} \label{sec:conclusion}
	We present results of a broadband X-ray study of Mrk\,926 using the 2009 \suzaku (Epoch 1), 2016 \xmm-\nustar\, (Epoch 2), and 2021 \nustar-\swift\textit{-XRT} (Epoch 3) observations. Our findings can be summarized as follows:
	\begin{enumerate}
		\item Mrk\,926 is a variable AGN. The unabsorbed 2-10\,keV luminosity of Mrk\,926 is 2.9$\pm{0.1}\times10^{44}$\,erg\,s$^{-1}$ (Epoch 1 and 2) and 1.1$\pm{0.1}\times10^{44}$\,erg\,s$^{-1}$ (Epoch 3).\\
		\item We found the broadband spectra of Mrk\,926 show a primary continuum, a soft excess and distant reflection. If the soft excess is modeled using relativistic reflection, we find some residuals above 30\,keV. We find no such residuals if we describe soft excess as a thermally Comptonized emission from a optically-thick warm corona.\\
		\item An optical Balmer satellite component was found in previous study of Mrk\,926 which suggests the presence of a compact jet. The signature of a compact jet in radio band is the presence of a flat/inverted spectrum that breaks at some higher frequency to a steep spectrum. We found a steep radio spectrum in the archival SED of Mrk\,926. Thus, the archival radio data cannot confirm the presence of a compact jet.
	\end{enumerate}
	Future multi-wavelength observing campaign including a high spatial resolution radio observation is highly desired to study the origin of this radio emitter in the `bare' Seyfert Mrk\,926.\\
	\section*{Acknowledgements}
	This work made use of data from the NuSTAR mission, a project led by the California Institute of Technology, managed by the Jet Propulsion Laboratory, and funded by the National Aeronautics and Space Administration. This research has also made use of the NASA/IPAC Extragalactic Database (NED), which is funded by the National Aeronautics and Space Administration and operated by the California Institute of Technology. We would like to thank Javier Garc{\'\i}a and Andy Fabian for helpful discussions.
	\section*{Data Availability}	
	All the data used in this article are publicly available from ESA \textit{XMM-Newton} Science Archive (XSA; http://nxsa.esac.esa.int/) and NASA High Energy Astrophysics Science Archive Research Center (HEASARC; https://heasarc.gsfc.nasa.gov/). The reduced data products used in this work may be shared on reasonable request to the authors.
	
	\bibliography{mendeley}{}

\begin{thebibliography}{}
\makeatletter
\relax
\def\mn@urlcharsother{\let\do\@makeother \do\$\do\&\do\#\do\^\do\_\do\%\do\~}
\def\mn@doi{\begingroup\mn@urlcharsother \@ifnextchar [ {\mn@doi@}
  {\mn@doi@[]}}
\def\mn@doi@[#1]#2{\def\@tempa{#1}\ifx\@tempa\@empty \href
  {http://dx.doi.org/#2} {doi:#2}\else \href {http://dx.doi.org/#2} {#1}\fi
  \endgroup}
\def\mn@eprint#1#2{\mn@eprint@#1:#2::\@nil}
\def\mn@eprint@arXiv#1{\href {http://arxiv.org/abs/#1} {{\tt arXiv:#1}}}
\def\mn@eprint@dblp#1{\href {http://dblp.uni-trier.de/rec/bibtex/#1.xml}
  {dblp:#1}}
\def\mn@eprint@#1:#2:#3:#4\@nil{\def\@tempa {#1}\def\@tempb {#2}\def\@tempc
  {#3}\ifx \@tempc \@empty \let \@tempc \@tempb \let \@tempb \@tempa \fi \ifx
  \@tempb \@empty \def\@tempb {arXiv}\fi \@ifundefined
  {mn@eprint@\@tempb}{\@tempb:\@tempc}{\expandafter \expandafter \csname
  mn@eprint@\@tempb\endcsname \expandafter{\@tempc}}}

\bibitem[\protect\citeauthoryear{Arnaud \& A.}{Arnaud \& A.}{1996}]{Arnaud1996}
Arnaud A. K.,  1996, Astronomical Data Analysis Software and Systems V, 101, 17

\bibitem[\protect\citeauthoryear{{Arnaud} et~al.,}{{Arnaud}
  et~al.}{1985}]{arnaud1985}
{Arnaud} K.~A.,  et~al., 1985, \mn@doi [\mnras] {10.1093/mnras/217.1.105},
  \href {https://ui.adsabs.harvard.edu/abs/1985MNRAS.217..105A} {217, 105}

\bibitem[\protect\citeauthoryear{{Ballantyne}}{{Ballantyne}}{2020}]{ballantyne2020a}
{Ballantyne} D.~R.,  2020, \mn@doi [\mnras] {10.1093/mnras/stz3294}, \href
  {https://ui.adsabs.harvard.edu/abs/2020MNRAS.491.3553B} {491, 3553}

\bibitem[\protect\citeauthoryear{{Ballantyne} \& {Xiang}}{{Ballantyne} \&
  {Xiang}}{2020}]{ballantyne2020b}
{Ballantyne} D.~R.,  {Xiang} X.,  2020, \mn@doi [\mnras]
  {10.1093/mnras/staa1866}, \href
  {https://ui.adsabs.harvard.edu/abs/2020MNRAS.496.4255B} {496, 4255}

\bibitem[\protect\citeauthoryear{{Balokovi{\'c}} et~al.,}{{Balokovi{\'c}}
  et~al.}{2018}]{borus2018}
{Balokovi{\'c}} M.,  et~al., 2018, \mn@doi [\apj] {10.3847/1538-4357/aaa7eb},
  \href {https://ui.adsabs.harvard.edu/abs/2018ApJ...854...42B} {854, 42}

\bibitem[\protect\citeauthoryear{{Bianchi}, {Matt}, {Balestra}, {Guainazzi}  \&
  {Perola}}{{Bianchi} et~al.}{2004}]{bianchi2004}
{Bianchi} S.,  {Matt} G.,  {Balestra} I.,  {Guainazzi} M.,   {Perola} G.~C.,
  2004, \mn@doi [\aap] {10.1051/0004-6361:20047128}, \href
  {https://ui.adsabs.harvard.edu/abs/2004A&A...422...65B} {422, 65}

\bibitem[\protect\citeauthoryear{{Blandford} \& {K{\"o}nigl}}{{Blandford} \&
  {K{\"o}nigl}}{1979}]{Blandford1979}
{Blandford} R.~D.,  {K{\"o}nigl} A.,  1979, \mn@doi [\apj] {10.1086/157262},
  \href {https://ui.adsabs.harvard.edu/abs/1979ApJ...232...34B} {232, 34}

\bibitem[\protect\citeauthoryear{{Choi}, {Dotani}, {Yi}, {Fletcher}  \&
  {Kim}}{{Choi} et~al.}{2001}]{choi2001}
{Choi} C.-S.,  {Dotani} T.,  {Yi} I.,  {Fletcher} A.,   {Kim} C.,  2001,
  \mn@doi [Journal of Korean Astronomical Society]
  {10.5303/JKAS.2001.34.3.129}, \href
  {https://ui.adsabs.harvard.edu/abs/2001JKAS...34..129C} {34, 129}

\bibitem[\protect\citeauthoryear{{Choi}, {Dotani}, {Chang}  \& {Yi}}{{Choi}
  et~al.}{2002}]{choi2002}
{Choi} C.-S.,  {Dotani} T.,  {Chang} H.-Y.,   {Yi} I.,  2002, \mn@doi [Journal
  of Korean Astronomical Society] {10.5303/JKAS.2002.35.1.001}, \href
  {https://ui.adsabs.harvard.edu/abs/2002JKAS...35....1C} {35, 1}

\bibitem[\protect\citeauthoryear{{Done}, {Davis}, {Jin}, {Blaes}  \&
  {Ward}}{{Done} et~al.}{2012}]{done2012}
{Done} C.,  {Davis} S.~W.,  {Jin} C.,  {Blaes} O.,   {Ward} M.,  2012, \mn@doi
  [\mnras] {10.1111/j.1365-2966.2011.19779.x}, \href
  {https://ui.adsabs.harvard.edu/abs/2012MNRAS.420.1848D} {420, 1848}

\bibitem[\protect\citeauthoryear{{Eckart}, {Witzel}, {Biermann}, {Johnston},
  {Simon}, {Schalinski}  \& {Kuhr}}{{Eckart} et~al.}{1986}]{eckart1986}
{Eckart} A.,  {Witzel} A.,  {Biermann} P.,  {Johnston} K.~J.,  {Simon} R.,
  {Schalinski} C.,   {Kuhr} H.,  1986, \aap, \href
  {https://ui.adsabs.harvard.edu/abs/1986A&A...168...17E} {168, 17}

\bibitem[\protect\citeauthoryear{{Evans} et~al.,}{{Evans}
  et~al.}{2009}]{evans2009}
{Evans} P.~A.,  et~al., 2009, \mn@doi [\mnras]
  {10.1111/j.1365-2966.2009.14913.x}, \href
  {https://ui.adsabs.harvard.edu/abs/2009MNRAS.397.1177E} {397, 1177}

\bibitem[\protect\citeauthoryear{{Fabian}, {Rees}, {Stella}  \&
  {White}}{{Fabian} et~al.}{1989}]{fabian1989}
{Fabian} A.~C.,  {Rees} M.~J.,  {Stella} L.,   {White} N.~E.,  1989, \mn@doi
  [\mnras] {10.1093/mnras/238.3.729}, \href
  {https://ui.adsabs.harvard.edu/abs/1989MNRAS.238..729F} {238, 729}

\bibitem[\protect\citeauthoryear{Garcia, Dauser, Reynolds, Kallman, McClintock,
  Wilms  \& Eikmann}{Garcia et~al.}{2013}]{Garcia2013}
Garcia J.,  Dauser T.,  Reynolds C.~S.,  Kallman T.~R.,  McClintock J.~E.,
  Wilms J.,   Eikmann W.,  2013, \mn@doi [The Astrophysical Journal]
  {10.1088/0004-637X/768/2/146}, 768, 146

\bibitem[\protect\citeauthoryear{Garcia et~al.,}{Garcia
  et~al.}{2014}]{Garcia2014}
Garcia J.,  et~al., 2014, \mn@doi [The Astrophysical Journal]
  {10.1088/0004-637X/782/2/76}, 782, 76

\bibitem[\protect\citeauthoryear{{Garc{\'\i}a} et~al.,}{{Garc{\'\i}a}
  et~al.}{2019}]{garcia2019}
{Garc{\'\i}a} J.~A.,  et~al., 2019, \mn@doi [\apj] {10.3847/1538-4357/aaf739},
  \href {https://ui.adsabs.harvard.edu/abs/2019ApJ...871...88G} {871, 88}

\bibitem[\protect\citeauthoryear{{George} \& {Fabian}}{{George} \&
  {Fabian}}{1991}]{george1991}
{George} I.~M.,  {Fabian} A.~C.,  1991, \mn@doi [\mnras]
  {10.1093/mnras/249.2.352}, \href
  {https://ui.adsabs.harvard.edu/abs/1991MNRAS.249..352G} {249, 352}

\bibitem[\protect\citeauthoryear{{Ghosh} \& {Soundararajaperumal}}{{Ghosh} \&
  {Soundararajaperumal}}{1992}]{ghosh1992}
{Ghosh} K.~K.,  {Soundararajaperumal} S.,  1992, \mn@doi [\apj]
  {10.1086/171873}, \href
  {https://ui.adsabs.harvard.edu/abs/1992ApJ...398..501G} {398, 501}

\bibitem[\protect\citeauthoryear{{Haardt} \& {Maraschi}}{{Haardt} \&
  {Maraschi}}{1991}]{haardt1991}
{Haardt} F.,  {Maraschi} L.,  1991, \mn@doi [\apjl] {10.1086/186171}, \href
  {https://ui.adsabs.harvard.edu/abs/1991ApJ...380L..51H} {380, L51}

\bibitem[\protect\citeauthoryear{{Haardt}, {Maraschi}  \&
  {Ghisellini}}{{Haardt} et~al.}{1994}]{haardt1994}
{Haardt} F.,  {Maraschi} L.,   {Ghisellini} G.,  1994, \mn@doi [\apjl]
  {10.1086/187520}, \href
  {https://ui.adsabs.harvard.edu/abs/1994ApJ...432L..95H} {432, L95}

\bibitem[\protect\citeauthoryear{Harrison et~al.,}{Harrison
  et~al.}{2013}]{Harrison2013}
Harrison F.~A.,  et~al., 2013, \mn@doi [The Astrophysical Journal]
  {10.1088/0004-637X/770/2/103}, 770, 103

\bibitem[\protect\citeauthoryear{{Jiang} et~al.,}{{Jiang}
  et~al.}{2019}]{jiang2019}
{Jiang} J.,  et~al., 2019, \mn@doi [\mnras] {10.1093/mnras/stz2326}, \href
  {https://ui.adsabs.harvard.edu/abs/2019MNRAS.489.3436J} {489, 3436}

\bibitem[\protect\citeauthoryear{Kalberla et~al.,}{Kalberla
  et~al.}{2005}]{Kalberla2005}
Kalberla P. M.~W.,  et~al., 2005, \mn@doi [Astronomy {\&} Astrophysics]
  {10.1051/0004-6361:20041864}, 440, 775

\bibitem[\protect\citeauthoryear{{Koljonen} et~al.,}{{Koljonen}
  et~al.}{2015}]{Koljonen2015}
{Koljonen} K.~I.~I.,  et~al., 2015, \mn@doi [\apj]
  {10.1088/0004-637X/814/2/139}, \href
  {https://ui.adsabs.harvard.edu/abs/2015ApJ...814..139K} {814, 139}

\bibitem[\protect\citeauthoryear{{Kollatschny} et~al.,}{{Kollatschny}
  et~al.}{2022}]{reverbmap}
{Kollatschny} W.,  et~al., 2022, \mn@doi [\aap] {10.1051/0004-6361/202142007},
  \href {https://ui.adsabs.harvard.edu/abs/2022A&A...657A.122K} {657, A122}

\bibitem[\protect\citeauthoryear{{Laha} \& {Ghosh}}{{Laha} \&
  {Ghosh}}{2021}]{laha2021}
{Laha} S.,  {Ghosh} R.,  2021, \mn@doi [\apj] {10.3847/1538-4357/abfc56}, \href
  {https://ui.adsabs.harvard.edu/abs/2021ApJ...915...93L} {915, 93}

\bibitem[\protect\citeauthoryear{{Mallick} et~al.,}{{Mallick}
  et~al.}{2018}]{mallick2018}
{Mallick} L.,  et~al., 2018, \mn@doi [\mnras] {10.1093/mnras/sty1487}, \href
  {https://ui.adsabs.harvard.edu/abs/2018MNRAS.479..615M} {479, 615}

\bibitem[\protect\citeauthoryear{{Matt}, {Perola}  \& {Piro}}{{Matt}
  et~al.}{1991}]{matt1991}
{Matt} G.,  {Perola} G.~C.,   {Piro} L.,  1991, \aap, \href
  {https://ui.adsabs.harvard.edu/abs/1991A&A...247...25M} {247, 25}

\bibitem[\protect\citeauthoryear{{McKaig}, {Ricci}, {Paltani}  \&
  {Satyapal}}{{McKaig} et~al.}{2022}]{polarsim}
{McKaig} J.,  {Ricci} C.,  {Paltani} S.,   {Satyapal} S.,  2022, \mn@doi
  [\mnras] {10.1093/mnras/stab3178}, \href
  {https://ui.adsabs.harvard.edu/abs/2022MNRAS.512.2961M} {512, 2961}

\bibitem[\protect\citeauthoryear{{Middei} et~al.,}{{Middei}
  et~al.}{2020}]{middei2020}
{Middei} R.,  et~al., 2020, \mn@doi [\aap] {10.1051/0004-6361/202038112}, \href
  {https://ui.adsabs.harvard.edu/abs/2020A&A...640A..99M} {640, A99}

\bibitem[\protect\citeauthoryear{{Mundell}, {Wilson}, {Ulvestad}  \&
  {Roy}}{{Mundell} et~al.}{2000}]{mundell2000}
{Mundell} C.~G.,  {Wilson} A.~S.,  {Ulvestad} J.~S.,   {Roy} A.~L.,  2000,
  \mn@doi [\apj] {10.1086/308318}, \href
  {https://ui.adsabs.harvard.edu/abs/2000ApJ...529..816M} {529, 816}

\bibitem[\protect\citeauthoryear{{Noda}, {Makishima}, {Nakazawa}, {Uchiyama},
  {Yamada}  \& {Sakurai}}{{Noda} et~al.}{2013}]{noda2013}
{Noda} H.,  {Makishima} K.,  {Nakazawa} K.,  {Uchiyama} H.,  {Yamada} S.,
  {Sakurai} S.,  2013, \mn@doi [\pasj] {10.1093/pasj/65.1.4}, \href
  {https://ui.adsabs.harvard.edu/abs/2013PASJ...65....4N} {65, 4}

\bibitem[\protect\citeauthoryear{{Panessa}, {Baldi}, {Laor}, {Padovani},
  {Behar}  \& {McHardy}}{{Panessa} et~al.}{2019}]{Panessa2019}
{Panessa} F.,  {Baldi} R.~D.,  {Laor} A.,  {Padovani} P.,  {Behar} E.,
  {McHardy} I.,  2019, \mn@doi [Nature Astronomy] {10.1038/s41550-019-0765-4},
  \href {https://ui.adsabs.harvard.edu/abs/2019NatAs...3..387P} {3, 387}

\bibitem[\protect\citeauthoryear{{Petrucci} et~al.,}{{Petrucci}
  et~al.}{2013}]{petrucci2013}
{Petrucci} P.~O.,  et~al., 2013, \mn@doi [\aap] {10.1051/0004-6361/201219956},
  \href {https://ui.adsabs.harvard.edu/abs/2013A&A...549A..73P} {549, A73}

\bibitem[\protect\citeauthoryear{{Petrucci}, {Ursini}, {De Rosa}, {Bianchi},
  {Cappi}, {Matt}, {Dadina}  \& {Malzac}}{{Petrucci}
  et~al.}{2018}]{petrucci2018}
{Petrucci} P.~O.,  {Ursini} F.,  {De Rosa} A.,  {Bianchi} S.,  {Cappi} M.,
  {Matt} G.,  {Dadina} M.,   {Malzac} J.,  2018, \mn@doi [\aap]
  {10.1051/0004-6361/201731580}, \href
  {https://ui.adsabs.harvard.edu/abs/2018A&A...611A..59P} {611, A59}

\bibitem[\protect\citeauthoryear{{Petrucci} et~al.,}{{Petrucci}
  et~al.}{2020}]{petrucci2020}
{Petrucci} P.~O.,  et~al., 2020, \mn@doi [\aap] {10.1051/0004-6361/201937011},
  \href {https://ui.adsabs.harvard.edu/abs/2020A&A...634A..85P} {634, A85}

\bibitem[\protect\citeauthoryear{{Piconcelli}, {Jimenez-Bail{\'o}n},
  {Guainazzi}, {Schartel}, {Rodr{\'\i}guez-Pascual}  \&
  {Santos-Lle{\'o}}}{{Piconcelli} et~al.}{2005}]{pico2005}
{Piconcelli} E.,  {Jimenez-Bail{\'o}n} E.,  {Guainazzi} M.,  {Schartel} N.,
  {Rodr{\'\i}guez-Pascual} P.~M.,   {Santos-Lle{\'o}} M.,  2005, \mn@doi [\aap]
  {10.1051/0004-6361:20041621}, \href
  {https://ui.adsabs.harvard.edu/abs/2005A&A...432...15P} {432, 15}

\bibitem[\protect\citeauthoryear{{Rivers}, {Markowitz}  \&
  {Rothschild}}{{Rivers} et~al.}{2011}]{rivers}
{Rivers} E.,  {Markowitz} A.,   {Rothschild} R.,  2011, \mn@doi [\apj]
  {10.1088/0004-637X/732/1/36}, \href
  {https://ui.adsabs.harvard.edu/abs/2011ApJ...732...36R} {732, 36}

\bibitem[\protect\citeauthoryear{{Ross}, {Fabian}  \& {Young}}{{Ross}
  et~al.}{1999}]{ross1999}
{Ross} R.~R.,  {Fabian} A.~C.,   {Young} A.~J.,  1999, \mn@doi [\mnras]
  {10.1046/j.1365-8711.1999.02528.x}, \href
  {https://ui.adsabs.harvard.edu/abs/1999MNRAS.306..461R} {306, 461}

\bibitem[\protect\citeauthoryear{{Smith} et~al.,}{{Smith} et~al.}{2020}]{jvla}
{Smith} K.~L.,  et~al., 2020, \mn@doi [\mnras] {10.1093/mnras/stz3608}, \href
  {https://ui.adsabs.harvard.edu/abs/2020MNRAS.492.4216S} {492, 4216}

\bibitem[\protect\citeauthoryear{{Ulvestad} \& {Wilson}}{{Ulvestad} \&
  {Wilson}}{1984}]{vla}
{Ulvestad} J.~S.,  {Wilson} A.~S.,  1984, \mn@doi [\apj] {10.1086/161821},
  \href {https://ui.adsabs.harvard.edu/abs/1984ApJ...278..544U} {278, 544}

\bibitem[\protect\citeauthoryear{{Walter} \& {Fink}}{{Walter} \&
  {Fink}}{1993}]{walter1993}
{Walter} R.,  {Fink} H.~H.,  1993, \aap, \href
  {https://ui.adsabs.harvard.edu/abs/1993A&A...274..105W} {274, 105}

\bibitem[\protect\citeauthoryear{{Weaver}, {Nousek}, {Yaqoob}, {Hayashida}  \&
  {Murakami}}{{Weaver} et~al.}{1995}]{weaver1995}
{Weaver} K.~A.,  {Nousek} J.,  {Yaqoob} T.,  {Hayashida} K.,   {Murakami} S.,
  1995, \mn@doi [\apj] {10.1086/176206}, \href
  {https://ui.adsabs.harvard.edu/abs/1995ApJ...451..147W} {451, 147}

\bibitem[\protect\citeauthoryear{{Weaver}, {Gelbord}  \& {Yaqoob}}{{Weaver}
  et~al.}{2001}]{weaver2001}
{Weaver} K.~A.,  {Gelbord} J.,   {Yaqoob} T.,  2001, \mn@doi [\apj]
  {10.1086/319713}, \href
  {https://ui.adsabs.harvard.edu/abs/2001ApJ...550..261W} {550, 261}

\bibitem[\protect\citeauthoryear{Wilms, Allen  \& McCray}{Wilms
  et~al.}{2000}]{Wilms2000}
Wilms J.,  Allen A.,   McCray R.,  2000, \mn@doi [The Astrophysical Journal]
  {10.1086/317016}, 542, 914

\bibitem[\protect\citeauthoryear{{Winter}, {Veilleux}, {McKernan}  \&
  {Kallman}}{{Winter} et~al.}{2012}]{winter2012}
{Winter} L.~M.,  {Veilleux} S.,  {McKernan} B.,   {Kallman} T.~R.,  2012,
  \mn@doi [\apj] {10.1088/0004-637X/745/2/107}, \href
  {https://ui.adsabs.harvard.edu/abs/2012ApJ...745..107W} {745, 107}

\bibitem[\protect\citeauthoryear{{{\.Z}ycki}, {Done}  \& {Smith}}{{{\.Z}ycki}
  et~al.}{1999}]{nthcomp}
{{\.Z}ycki} P.~T.,  {Done} C.,   {Smith} D.~A.,  1999, \mn@doi [\mnras]
  {10.1046/j.1365-8711.1999.02885.x}, \href
  {https://ui.adsabs.harvard.edu/abs/1999MNRAS.309..561Z} {309, 561}

\makeatother
\end{thebibliography}
	\bibliographystyle{mnras}
	\bsp	
	\label{lastpage}	
\end{document}